\documentstyle[12pt]{article}
\makeatletter
\@addtoreset{equation}{section}
\makeatother

\begin{document}
\begin{center}
{\Large \bf
Another Comment on ``Quantum Entanglement and the
Nonexistence of Superluminal Signals''} \\[1.5cm]
{\bf Vladimir S.~MASHKEVICH}\footnote {E-mail:
mash@gluk.apc.org}  \\[1.4cm]
{\it Institute of Physics, National academy
of sciences of Ukraine \\
252028 Kiev, Ukraine} \\[1.4cm]
\vskip 1cm

{\large \bf Abstract}
\end{center}

We argue that the claim given in quant-ph/9801014 remains
untenable in the revised version. The fallacy in the proof
is a misinterpretation of the no-cloning and teleportation
theorems, which do not involve time and reference frames.

\newpage

\section*{\Large \bf Introduction}

In a recent paper [1] Westmoreland and Schumacher made an
attempt to show that ``ordinary quantum mechanics is not
consistent with the superluminal transmission of classical
information''. Their proof was constructed from three
elements: the no-cloning theorem, quantum teleportation,
and the relativity of simultaneity.

In a comment [2] we argued that the claim given in [1] was
untenable, since the formulation of the no-cloning theorem
did not allow quantum jumps.

Now Westmoreland and Schumacher have replaced [1] with a
revised version. Although the proof of the no-cloning
theorem has remained the same, the new version includes
an essential note: ``...the `no-cloning' theorem in fact
holds for the most general sort of quantum evolution
described by a completely positive map on density
operators... In particular, cloning is still impossible
even if we allow measurements and manipulations of the systems
based on the outcomes of measurements.''

But using the extended version of the no-cloning theorem does
not save the proof given in [1]. The fallacy is a
misinterpretation of the no-cloning and teleportation theorems,
which do not involve time and reference frames. Indeed,
signals---superluminal or not---imply no contradiction between
teleportation and the no-cloning theorem: the former is possible
and the latter is valid.

\section{The no-cloning theorem}

Let $T$ be a completely positive map on the set of the states,
i.e., statistical operators $\rho$ of a system which is of the
following form:
\begin{equation}
T\rho=\sum_{i}V_{i}\rho V^{\dagger}_{i},\qquad V_{i}=
U_{i}P_{i}\;\;\;{\rm or}\;\;\;P_{i}U_{i},
\label{1.1}
\end{equation}
where $U_{i}$ is a unitary operator, $P_{i}$ is a projector, and
\begin{equation}
\sum_{i}P_{i}=I.
\label{1.2}
\end{equation}
Let us consider a composed $CBA$ system, the Hilbert spaces of
$B$ and $C$ systems being identified. We use the notations:
\begin{equation}
\rho=\rho_{CBA},\;\;\rho_{C}={\rm Tr}_{BA}\rho,\;\;
\rho_{BA}={\rm Tr}_{C}\rho,\;\;(T\rho)_{C}={\rm Tr}_{BA}(T\rho),
\;\;{\rm etc}.
\label{1.3}
\end{equation}

{\it The no-cloning theorem\/}:
\begin{equation}
(\forall(T,\rho_{BA}))(\exists\rho_{C}):(T\rho)_{B}\ne\rho_{C}
\;\;{\rm or}\;\;(T\rho)_{C}\ne\rho_{C},
\label{1.4}
\end{equation}
or, equivalently,
\begin{equation}
(\neg\exists(T,\rho_{BA}))(\forall\rho_{C}):(T\rho)_{B}=\rho_{C}
\;\;{\rm and}\;\;(T\rho)_{C}=\rho_{C}.
\label{1.5}
\end{equation}
In words: There do not exist $T$ and $\rho_{BA}$, such that for
every $\rho_{C}$
\begin{equation}
(T\rho)_{B}=(T\rho)_{C}=\rho_{C}
\label{1.6}
\end{equation}
holds.

\section{Teleportation}

The essence of teleportation may be formulated as

{\it The teleportation theorem\/}:
\begin{equation}
(\exists\;CBA\;{\rm system})(\exists(T,\rho_{BA}))
(\forall\rho_{C}):(T\rho)_{B}=\rho_{C}.
\label{2.1}
\end{equation}
In words: There exist $CBA$ system, $T$, and $\rho_{BA}$, such
that for every $\rho_{C}$
\begin{equation}
(T\rho)_{B}=\rho_{C}
\label{2.2}
\end{equation}
holds.

{\it Corollary\/}: For every teleportation
\begin{equation}
(T\rho)_{C}\ne\rho_{C}
\label{2.3}
\end{equation}
holds (no cloning [3]).

\section{What is the proof}

Let $P_{\psi}$ be a projector corresponding to a vector $\psi$,
$\|\psi\|=1$. In [1]
\begin{equation}
\rho=\rho_{C}\otimes\rho_{BA},\quad \rho_{C}=P_{\phi_{C}},\quad
\rho_{BA}=P_{\Psi^{-}_{AB}},
\label{3.1}
\end{equation}
the teleportation map is
\begin{equation}
T\rho=\sum_{i=1}^{4}V_{i}\rho V_{i}^{\dagger},\quad
V_{i}=U_{Bi}\otimes P_{ACi},\quad
P_{ACi}=P_{\chi_{ACi}},
\label{3.2}
\end{equation}
where
\begin{equation}
\chi_{AC1}=\Psi^{+}_{AC},\quad \chi_{AC2}=\Psi^{-}_{AC},
\quad \chi_{AC3}=\Phi^{+}_{AC},\quad \chi_{AC4}=\Phi^{-}_{AC}.
\label{3.3}
\end{equation}

The proof is as follows. Due to superluminal signals, there
exists a frame of reference in which for any time $t$ such that
\begin{equation}
t_{{\rm II}}<t<t_{{\rm I}}
\label{3.4}
\end{equation}
the states of $B$ and $C$ systems are
\begin{equation}
\rho^{B}_{t}=(T\rho)_{B}=\rho_{C}
\label{3.5}
\end{equation}
and
\begin{equation}
\rho^{C}_{t}=\rho_{C}
\label{3.6}
\end{equation}
respectively, so that
\begin{equation}
\rho^{B}_{t}=\rho^{C}_{t}=\rho_{C}\qquad {\rm for}\quad t\in
(t_{{\rm I}},t_{{\rm II}}).
\label{3.7}
\end{equation}
Eq.(\ref{3.7}), the authors conclude, contradicts the no-cloning
theorem, which completes the proof.

\section{What is wrong}

The no-cloning and teleportation theorems do not involve
time and reference frames. Therefore the only relation
in the proof which is connected with the theorems is
\begin{equation}
(T\rho)_{B}=\rho_{C}.
\label{4.1}
\end{equation}
A contradiction would be the equality
\begin{equation}
(T\rho)_{C}=\rho_{C},
\label{4.2}
\end{equation}
which  does not hold.

As for signals---superluminal or not---their only function
is to coordinate sets
\begin{equation}
\{P_{ACi}\}_{i=1}^{4}\quad {\rm and}\quad \{U_{Bi}\}_{i=1}^{4}.
\label{4.3}
\end{equation}

\section*{Acknowledgment}

I would like to thank Stefan V. Mashkevich for helpful
discussions.

\end{document}